\documentclass[%
 reprint,
 amsmath,amssymb,
 aps,
]{revtex4-2}

\usepackage{graphicx}
\usepackage{dcolumn}
\usepackage{bm}

\begin{document}

\preprint{APS/123-QED}

\title{Anomalous cooling and overcooling of active systems}

\author{Fabian Jan Schwarzendahl}\email{Fabian.Schwarzendahl@hhu.de}
\affiliation{Institut f\"ur Theoretische Physik II: Weiche Materie, Heinrich-Heine-Universit\"at D\"usseldorf, 40225 D\"usseldorf, Germany
}
\author{Hartmut L\"owen}%
\affiliation{Institut f\"ur Theoretische Physik II: Weiche Materie, Heinrich-Heine-Universit\"at D\"usseldorf, 40225 D\"usseldorf, Germany
}
\date{\today}

\begin{abstract}
The phenomenon that a system at a hot temperature cools faster than at a warm temperature, referred to as the Mpemba effect, has been recently realized for trapped colloids. Here, we investigate the cooling and heating process of a self-propelling active colloid using numerical simulations and theoretical calculations with a model that can directly be tested in experiments. Upon cooling the particles' active motion induces a Mpemba effect. 
Transiently the system can even exhibit smaller temperatures than its final temperature, a surprising phenomenon which we refer to as activity-induced overcooling. 
\end{abstract}

\maketitle
When water is cooled down to be frozen, it seems intuitive that the cooler the water the faster it will freeze. Contrary to that, about 2300 years ago, Aristotle already noticed that "to cool hot water quickly, begin by putting it in the sun"~\cite{aristotle}, observing that hot water can be cooled and also frozen faster than warm water. The first systematic study to investigate this effect was conducted in the 1960s by Mpemba~\cite{mpemba1969cool}, after whom it was named.
Thereafter, numerous experimental studies followed but no consensus on the cause of the Mpemba effect for water was found so far~\cite{jeng2006mpemba,wojciechowski1988freezing,auerbach1995supercooling,vynnycky2012axisymmetric,vynnycky2015can,burridge2016questioning}.
Recently, the Mpemba effect was discovered for colloidal particles that are subjected to a thermal quench~\cite{kumar2020exponentially}, where the colloids were confined to a double well potential mimicking the liquid and frozen state of water.
The experimental findings match theoretical predictions giving a clear explanation of the underlying effect~\cite{bechhoefer2021fresh,chetrite2021metastable} and provide an experimental road to recent theoretical advances in understanding the Mpemba effect~\cite{lu2017nonequilibrium,klich2019mpemba,gal2020precooling,carollo2021exponentially}.
\begin{figure}
    \centering
    \includegraphics[width=1.0\columnwidth]{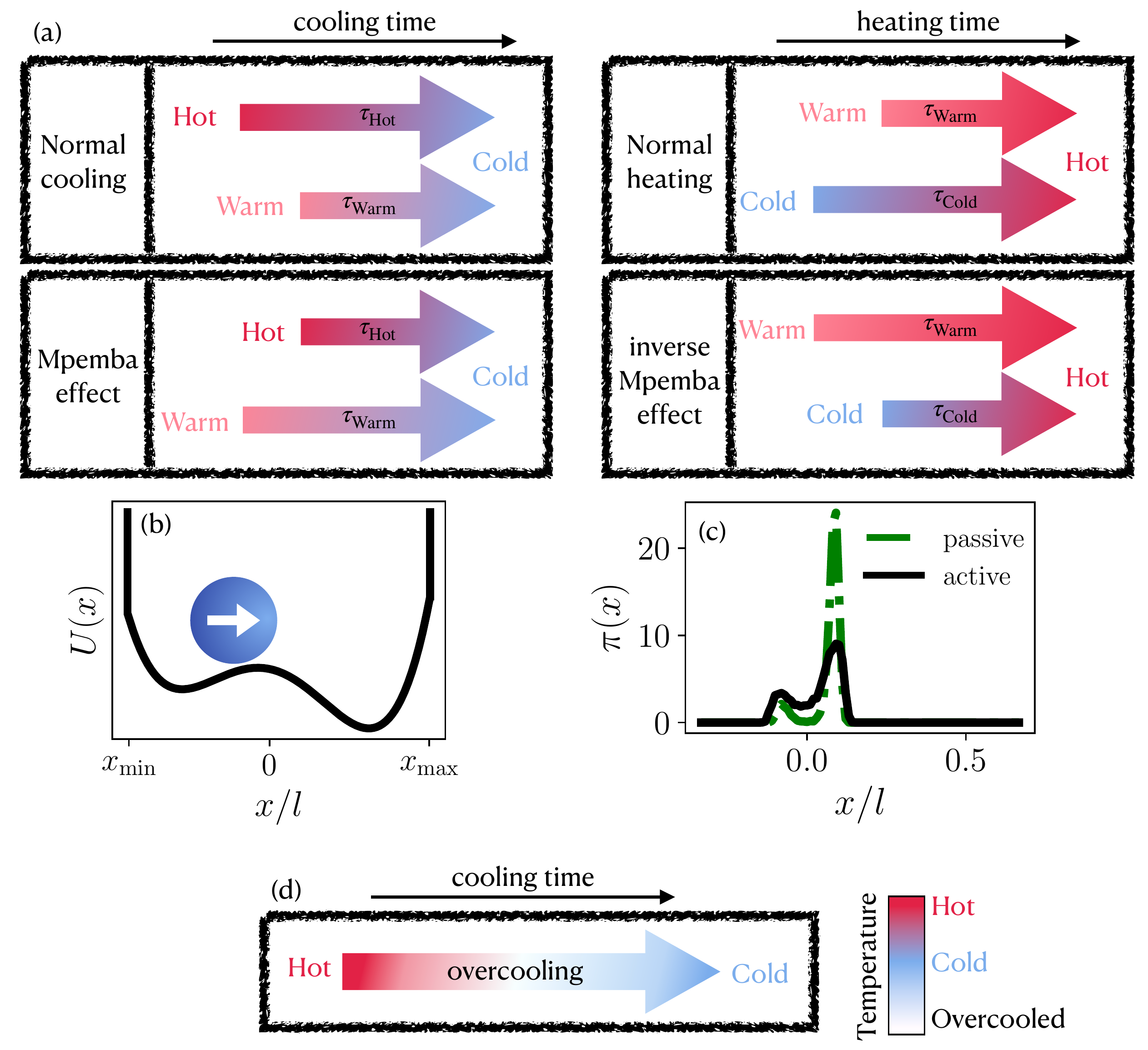}
    \caption{ (a) Cooling and heating scenarios, where each colored arrow represents a cooling/heating process. The arrows' length displays the time that the system needs to cool down or heat up. (b) Asymmetric potential with walls $U(x)$ as function of space that the active colloid (blue circle) is subjected to. (c) Occupation probability $P(x)$ of an active (solid black line) and passive (dashed-dotted green line) colloid in the external potential shown in (b).
    (d) An overcooling scenario, where during the cooling process lower temperatures than the final temperature are reached. The arrow's color code represents the temperature.}
    \label{fig:Mpemba_sketch_pot_dist}
\end{figure}

Cooling of a liquid or the thermal quench of a colloidal particle is a nonequilibrium process since decreasing the temperature effectively removes energy from the system. Nonetheless, the constituents involved in the cooling process, i.e. the liquid or colloids, are passive and do not themselves add or remove energy from the system. In contrast to that, active colloids that are self-propelling, constantly pump energy into the system and are therefore inherently out of equilibrium~\cite{bechinger2016active,gompper20202020}. Active colloidal particles have been realized in various experimental systems~\cite{bechinger2016active} and can show fascinating effects such as wall accumulation~\cite{narinder2019active,volpe2011microswimmers,maggi2016self}, activity induced ratchet motion~\cite{di2010bacterial,rodenburg2018ratchet}, emergent nonequilibrium fluxes~\cite{cammann2021emergent}, motility induced phase separation~\cite{buttinoni2013dynamical,palacci2013living,bialke2012crystallization}, or vortices~\cite{bricard2015emergent}.
Given their nonequilibrium nature, we ask here if and under which conditions active colloids can exhibit a Mpemba effect and if activity can change it. Surprisingly, we find that activity induces an \textit{overcooling} of the system, where it transiently reaches a temperature that is lower than its final steady state temperature (Fig.~\ref{fig:Mpemba_sketch_pot_dist}(d)).

\begin{figure}[!t]
    \centering
    \includegraphics[width=1.0\columnwidth]{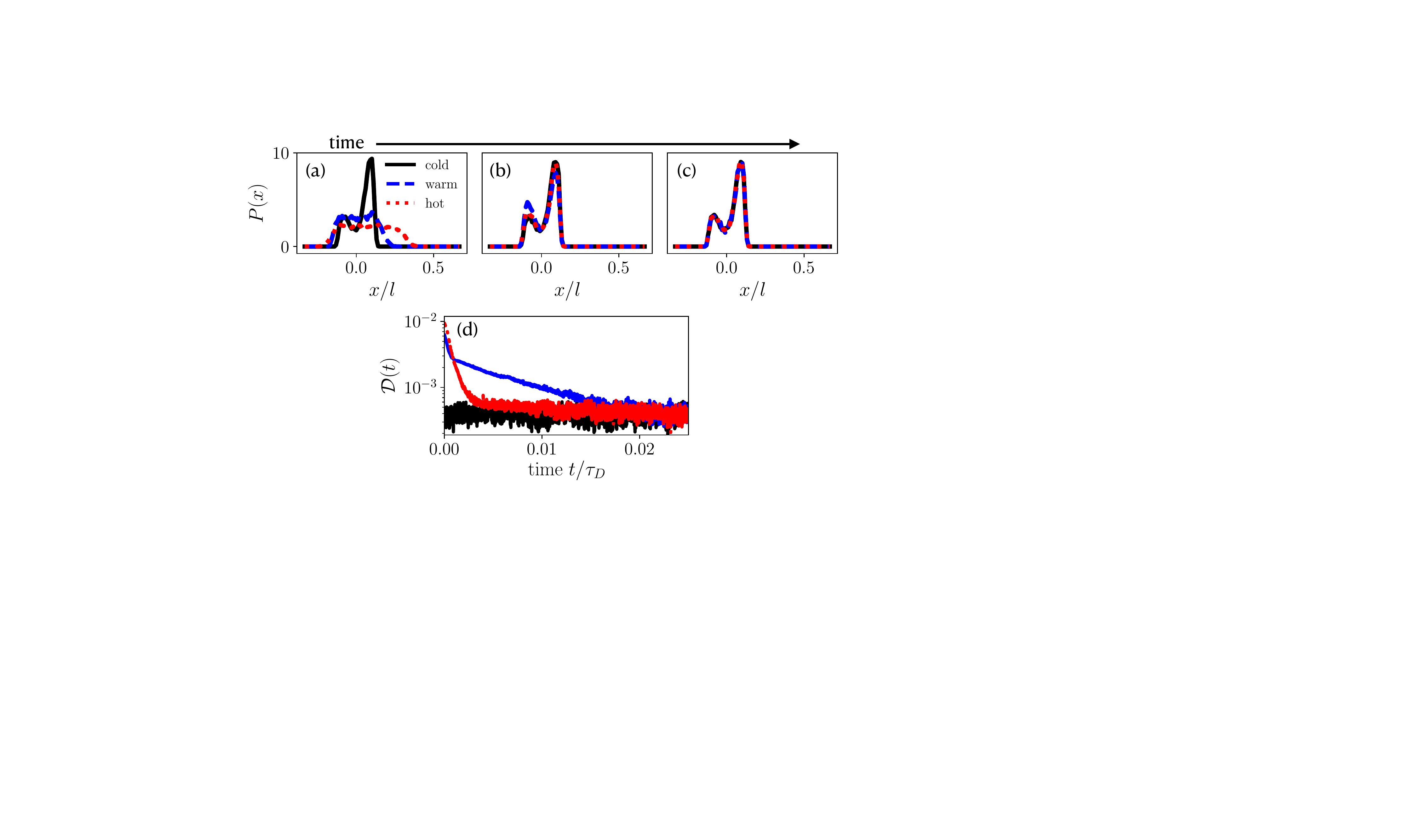}
    \caption{Probability distributions of cooling active colloids. We show a hot (red dotted line, $T_{\mathrm{hot}}=2500 T_{\mathrm{cold}}$), a warm (blue dashed line, $T_{\mathrm{warm}}=50 T_{\mathrm{cold}}$) and cold (black solid line) colloid. 
    (a) Initial steady state distributions at time $t=0$. A temperature quench is applied at this time. (b) Distributions at time $t=3\times 10^{-3} \tau_D$. The hot colloid has relaxed to $T_{\mathrm{cold}}$. (c) Distributions at time $t=50 \times 10^{-3} \tau_D$. The warm colloid has relaxed to $T_{\mathrm{cold}}$. (d) Distance $\mathcal{D}(t)$ between probability distributions for a hot (red line), warm (blue line), and cold (black line) colloid, compared to  the steady state of a cold colloid. }
    \label{fig:cooling_dists}
\end{figure}

To illustrate the Mpemba effect, we imagine two systems, one of which is at an initial warm temperature and the second is at an initial hot temperature and ask the systems to cool down to an imposed cold temperature (Fig.~\ref{fig:Mpemba_sketch_pot_dist}(a)).
Normally, the warm system cools faster than the hot system. A Mpemba effect occurs if hot system cools faster than the warm system.
Analogously, a cold system can heat up faster than the warm system (Fig.~\ref{fig:Mpemba_sketch_pot_dist}(a)), which is the inverse Mpemba efffect~\cite{lu2017nonequilibrium,kumar2021anomalous}.



To identify scenarios like that we investigate the cooling and heating process of active colloids in a confining asymmetric potential (Fig.~\ref{fig:Mpemba_sketch_pot_dist}(b)). We model the colloid as an active Brownian particle and show that it exhibits anomalous cooling, which is induced by its active motion. These results are supported using theoretical calculations based on Master equations, which give further insight into the relaxation process described by a series of exponential decays. 

We consider an active Brownian particle (ABP) modeling a colloid that is suspended in a bath at temperature $T$. We assume an overdamped motion of the particle in one spacial dimension that is described by the following equation
\begin{align}
    \frac{\mathrm{d} x}{\mathrm{d} t} 
    = v_0 n - \partial_x U(x) + \eta,
    \label{eq:ABP}
\end{align}
where $v_0$ is the particles self-propulsion speed, and $n$ is the direction of propulsion. The particle's propulsion direction is inverted after a time $t_p$, which follows an exponential distribution $p(t_p)= 1/\tau_{p} e^{t_p/\tau_{p}}$ with a persistence time $\tau_{p}$.
Furthermore, $\eta$ is a Gaussian white noise with zero mean and variance $\langle \eta(t) \eta(t') \rangle =2 D_T \delta(t-t')$, where $D_T$ is the particle's translational diffusion constant. The latter is controlled by the temperature $T$ of the bath, that is $D_T= \frac{k_B T}{\gamma}$, where $k_B$ is the Boltzmann constant, and $\gamma$ is the particle's friction coefficient. The particle is exposed to an external  double well potential $U(x)$ whose explicit form is
\begin{align}
U(x) = \begin{cases}
    -F_0 x, & \text{if $x<x_{\mathrm{min}}$},\\
    F_B \left( (1-x^2)^2 - \frac{1}{2} x \right), & \text{if $x_{\mathrm{min}}<x<x_{\mathrm{max}}$},\\
    F_0 x, & \text{if $x>x_{\mathrm{max}}$}.\\
  \end{cases}
  \label{eq:ext_pot}
\end{align}
The potential Eq.~\eqref{eq:ext_pot} is displayed in Fig.~\ref{fig:Mpemba_sketch_pot_dist}(b), where the terms in Eq.~\eqref{eq:ext_pot} proportional to $F_0$ model repulsive walls at positions $x_{\mathrm{min}}$ and $x_{\mathrm{max}}$ and the term proportional to $F_B$ models an asymmetric potential with two minima of different height. The size of the box to which our particle is confined $l= |x_{\mathrm{max}}-x_{\mathrm{min}}|$ gives a natural unit of length and together with the translational diffusion constant, we find a natural unit of time $\tau_D= l^2/D_T$ and velocity $v_D=l/\tau_D$. Experimentally the system described by Eq.~\eqref{eq:ABP} can for example be realized using a Janus colloid~\cite{bechinger2016active,gomez2017tuning}, yielding a self-propulsion together with optical~\cite{shen2019far,kumar2020exponentially} or acoustic~\cite{takatori2016acoustic} traps that establish an external potential. Similar models for active Brownian particles in a double well potential have been considered in \cite{caprini2021correlated,caprini2019active,sharma2017escape,woillez2020nonlocal,fily2019self,woillez2019activated,zanovello2021optimal,scacchi2018mean}.

In the following we simulate many realizations of Eq.~\eqref{eq:ABP} which yield the probability distribution $P(x,t)$ to find the particle at position $x$ and time $t$. Typical steady state distributions denoted by $\pi(x)$ for a passive ($v_0=0$) and an active particle ($v_0\neq0$) are shown in Fig.~\ref{fig:Mpemba_sketch_pot_dist}(c), where the particles are localised around the two minima of the external potential. Additionally, the active particles' probability distribution is enhanced towards the wall regions due to the well known wall accumulation effect that arises from to the particle's persistence and active motion~\cite{elgeti2015physics,elgeti2009self,elgeti2013wall,kaiser2012capture,ostapenko2018curvature,angelani2017confined,elgeti2015run,tailleur2009sedimentation,wittmann2016active,schaar2015detention,malgaretti2017model}.

\begin{figure}[!t]
    \centering
    \includegraphics[width=1.0\columnwidth]{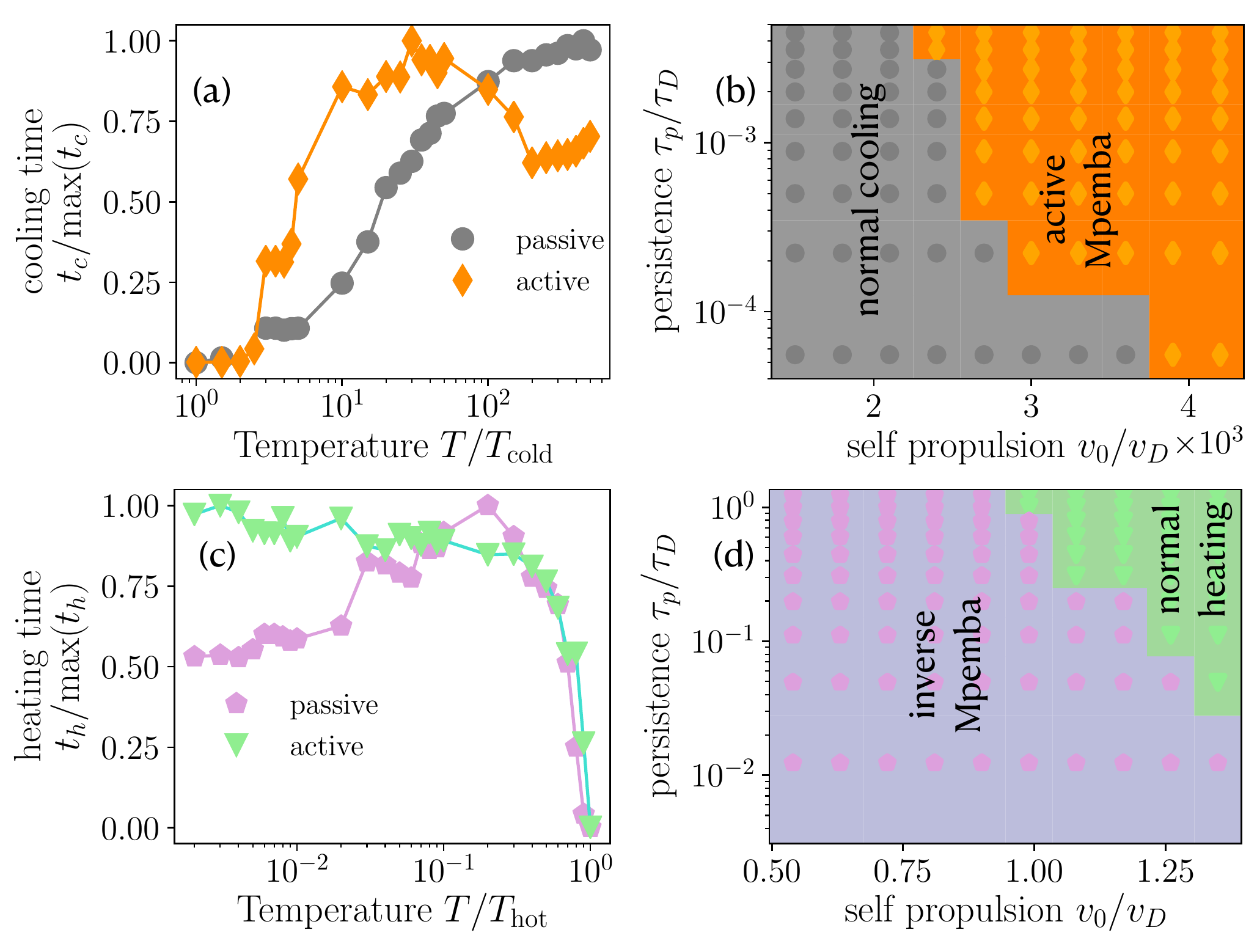}
    \caption{(a) Cooling time as function of initial temperature $T$ to final temperature $T_{\mathrm{cold}}$ for passive (gray circles) and active (orange diamonds) particles ($v_0=3.9\times 10^3 v_D$, $\tau_p=5\times 10^{-4} \tau_D$). (b) Cooling phase diagram as function of self propulsion and persistence showing normal cooling and the active Mpemba effect. (c) Heating time as function of initial temperatures $T$ to final temperature $T_{\mathrm{hot}}$ for active (green triangles, $v_0=9 v_D$, $\tau_p= 1.2\tau_D$) and passive (purple pentagons) particle. (d) Heating phase diagram for varying self propulsion and persistence showing the inverse Mpemba effect and its elimination as normal heating.}
    \label{fig:phase_diags}
\end{figure}

Let us now consider two active Brownian particle systems, which are in their nonequilibrium steady state at a hot and warm temperature (Fig.~\ref{fig:cooling_dists}(a)). The respective temperatures are $T_{\mathrm{hot}}$ and $T_{\mathrm{warm}}$ with $T_{\mathrm{hot}} >T_{\mathrm{warm}}$.
We then apply a thermal quench to each particle, that is we instantaneously change each particle's temperature to $T_{\mathrm{cold}}$.
Figure~\ref{fig:cooling_dists}(a)-(c) show how the distributions of the warm and hot particle relax and finally reach their new steady state distributions $\pi_{\mathrm{cold}}(x)$ at temperature $T_{\mathrm{cold}}$ (Fig.~\ref{fig:cooling_dists}(c)). Intriguingly, we find the hot particle relaxes faster (Fig.~\ref{fig:cooling_dists}(b)) than the warm particle (Fig.~\ref{fig:cooling_dists}(c)). Hence, we observe that the hot particle cools faster than the warm particle, which is an anomalous cooling, i.e. a Mpemba effect. Initially the warm particles' distribution has a stronger localization into the minima of the external potential Eq.~\eqref{eq:ext_pot} than the hot particle (Fig.~\ref{fig:cooling_dists} (a)).
This means that the effective barrier that a warm particle has to overcome to relax to a cold temperature is higher than the barrier of the hot particle. Therefore, the hot particle can cool faster than the warm particle.

To get further insight into the relaxation process, we study how far apart the cooled steady state distribution $\pi_{\mathrm{cold}}(x)$ and the probability distribution $P(x,t)$  of a particle are from each other during a cooling process. Explicitly, we discretize the spatial components of both $\pi_{\mathrm{cold}}(x)$ and $P(x,t)$ into $N$ grid points, giving $\pi_{i,\mathrm{cold}}$ and $P_i(t)$ respectively and compute the distance measure
\begin{align}
    \mathcal{D}(t) = \sum_{i=0}^N \left| P_i(t) - \pi_{i,\mathrm{cold}} \right|.
    \label{eq:dist_measure}
\end{align}
 Figure~\ref{fig:cooling_dists}(d) shows the distance measure Eq.~\eqref{eq:dist_measure} which quantifies the cooling process of the hot and warm particle, again showing that the hot particle relaxes and therefore cools faster than the cold particle. From this measure we can extract a cooling time $t_c$, defined as the time at which $\mathcal{D}(t)$ has decayed to zero, or here to the noise level.

\begin{figure}[!t]
    \centering
    \includegraphics[width=1.0\columnwidth]{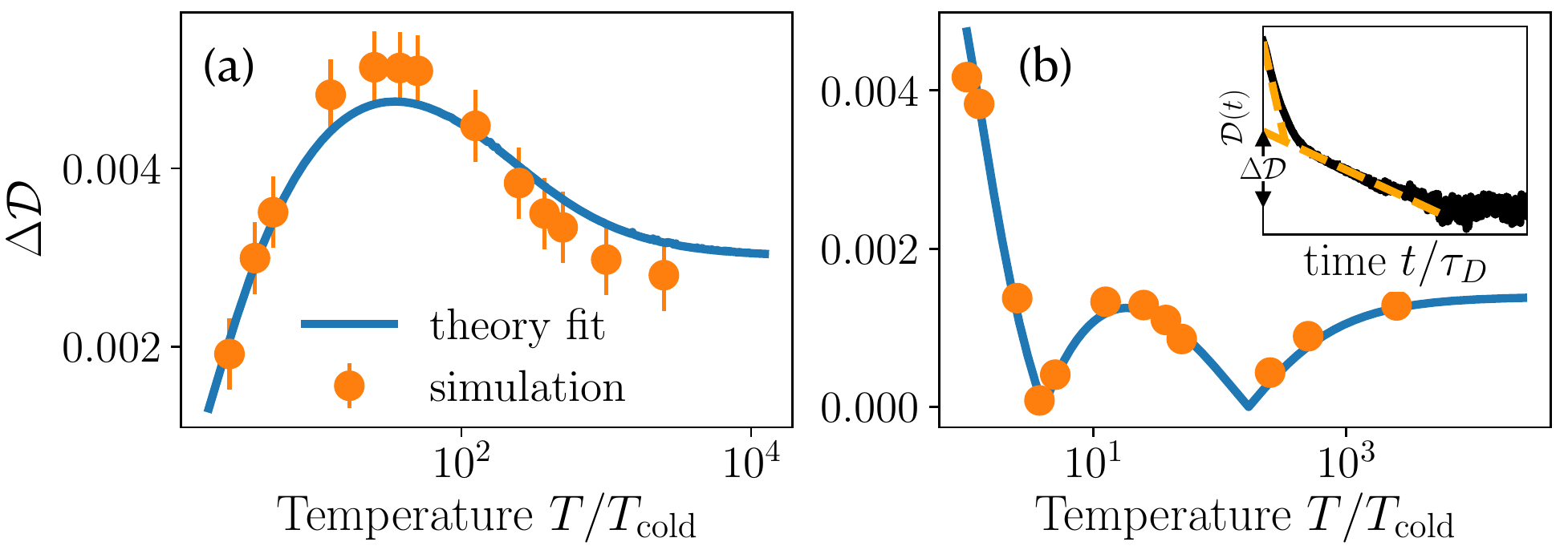}
    \caption{ Cooling process of temperatures $T$ to $T_{\mathrm{cold}}$, quantified by $\Delta \mathcal{D}$. Solid blue line shows theoretical calculation and orange data points are extracted from our numerical approach (inset (b)). In (a) the initial state is active whereas in (b) the initial state is passive. (b) (inset) Example cooling process (black solid line) on a half logarithmic scale, to which we fit two exponential functions (orange dashed lines) to obtain $\Delta \mathcal{D}$ (see ordinate).
    }
    \label{fig:comparison_theory}
\end{figure}

Using the cooling time $t_c$ we explore a range of initial temperatures $T$, each of which we cool down to $T_{\mathrm{cold}}$, yielding cooling curves as shown in Fig.~\ref{fig:phase_diags}(a). For passive particles ($v_0=0$) our system shows normal cooling (Fig.~\ref{fig:phase_diags}(a)), that is the cooling time increases monotonously with initial temperature. Contrary to that if we turn on the active motion ($v_0\neq 0$), the cooling curve becomes nonmonotonous, i.e. a hot temperature cools faster than a cold one, which is a Mpemba effect that is induced by activity. 
Here, activity changes the probability distributions of the particles by inducing a wall accumulation (Fig.~\ref{fig:Mpemba_sketch_pot_dist}(c)), which in turn enables the Mpemba effect.
To scrutinize the dependence of the active Mpemba effect on the particle's self propulsion and persistence, we build a phase diagram shown in Fig~\ref{fig:phase_diags}(b) (see SI for details). As expected, for low persistence and self propulsion we have a situation that is similar to a passive particle and no Mpemba effect is found. Increasing both the self propulsion and persistence then leads to an active Mpemba effect.

\begin{figure*}
    \centering
    \includegraphics[width=2.0\columnwidth]{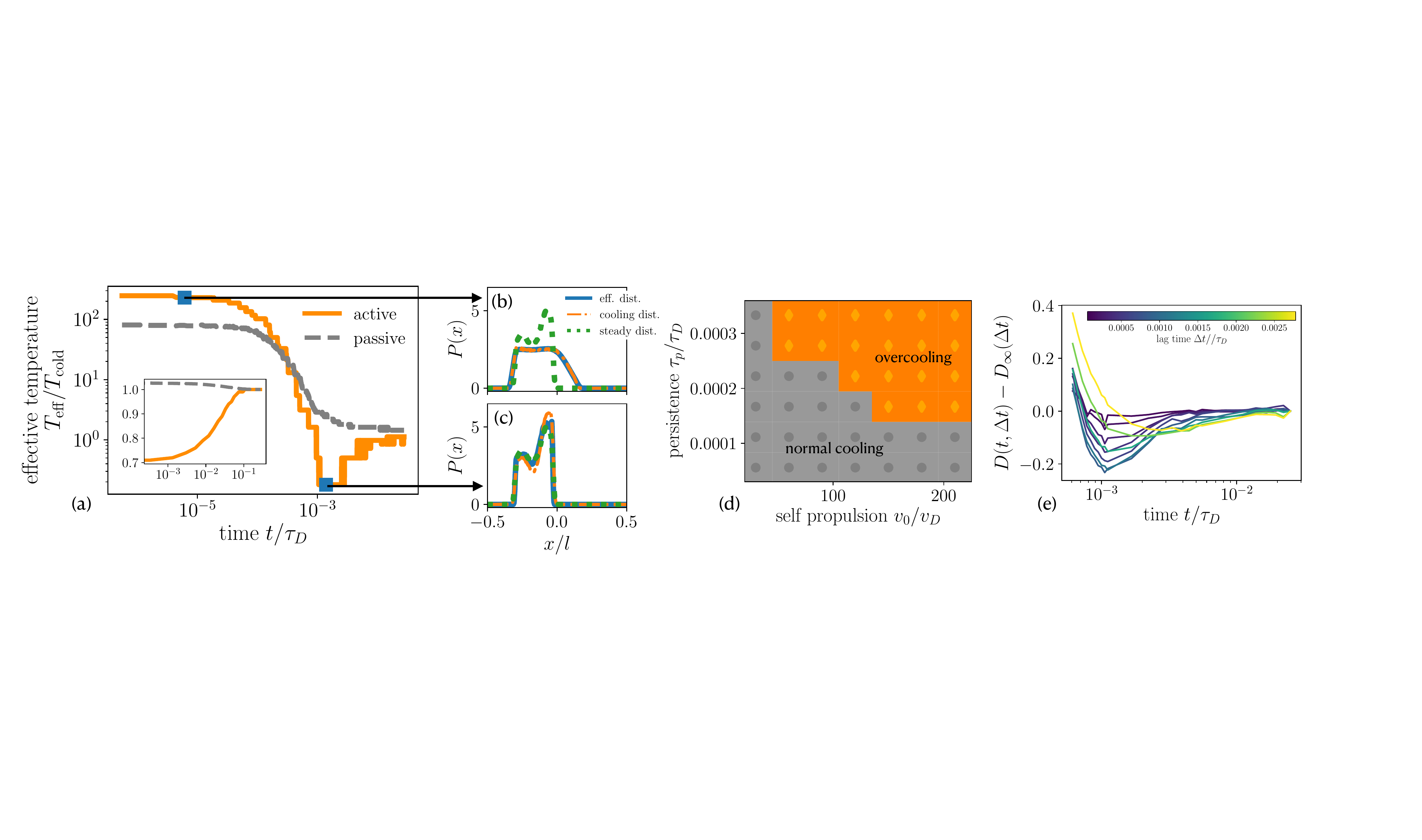}
    \caption{
    (a) Effective temperature $T_{\mathrm{eff}}$ as function of time for an active (orange solid line) and passive (grey dashed line). (inset) Effective temperature $T_{\mathrm{eff}}$  extracted from our theoretical approach.
    (b)-(c) Probability distributions of the cooling process (orange dashed dotted line), effective distribution $\pi_{T_{\mathrm{eff}}}$  (solid blue line) and final steady state (green dotted line). (b) Shows a typical state at the beginning of the cooling process  and (c) a state where the system is overcooled.
    (d) Phase diagram showing overcooling (orange diamonds) and normal cooling (grey cricles) for varying self-propulsion and persistence.
    (e) Reduced lag diffusion as a function of time for different lag times (color code).}
    \label{fig:overcooling}
\end{figure*}

Analogous to cooling we study the heating process of a particle towards the temperature $T_\mathrm{hot}$.  Here, our passive particle has a nonmonotonous heating curve (Fig.~\ref{fig:phase_diags}(c)), an inverse Mpemba effect, which is eliminated by the active motion. A similar picture arises in the phase diagram (Fig.~\ref{fig:phase_diags}(d)), where in the limit of low self propulsion and persistence, we find an inverse Mpemba effect, which is vanishes for increasing self propulsion and persistence.

In order to get a deeper insight into the relaxation process of our active Brownian particle we move to a statistically equivalent version of Eq.~\eqref{eq:ABP} in terms of a probability $P(x,t)$ to find the particle at position $x$ and time $t$ and its polarization $P^*(x,t)$. Explicitly we have
\begin{align}
    \partial_t P &= v_0 \partial_x P^* - \partial_x \left[ (\partial_x U(x)) P \right] + D_T \partial_x^2 P
    \label{eq:probability} \\
        \partial_t P^* &= v_0 \partial_x P - \partial_x \left[ (\partial_x U(x)) P^* \right] + D_T \partial_x^2 P^* - \frac{2}{\tau_p} P^*,
    \label{eq:polarization}
\end{align}
where the first terms on the right hand side of both Eq.~\eqref{eq:probability}-\eqref{eq:polarization} stem from the self propulsion of the active particle, the second terms stem from the external  potential (Eq.~\eqref{eq:ext_pot}) and the third terms account for translational diffusion, which in turn is controlled by the temperature, as described above. The last term in Eq.~\eqref{eq:polarization} comes  from the persistent motion of the particle, which is turned around after a persistence time $\tau_p$ (for a derivation of Eq.~\eqref{eq:probability}-\eqref{eq:polarization} see SI).

Relating to our cooling process using a thermal quench, we have to solve Eq.~\eqref{eq:probability} with temperature $T_{\mathrm{cold}}$, and an initial distribution $P(x,0)= \pi_{\mathrm{ini}}(x)$ that has a hot/warm initial temperature~\cite{lu2017nonequilibrium}. This problem can be tackled using an eigenfunction expansion~\cite{risken1996fokker}, giving the formal solution
\begin{align}
    P(x,t) = \pi_\mathrm{cold}(x) + \sum_{n=2}^{\infty} a_{n, \mathrm{ini}}
 e^{-\lambda_n t} v_n(x),
 \label{eq:eigenfunction_expansion}
\end{align}
where $v_n(x)$ are the eigenfunctions of Eq.~\eqref{eq:probability}, $\lambda_n$ are the eigenvalues, and $a_{n, \mathrm{ini}}$ are the overlap values with the initial condition (for details see SI). Here, the eigenvalues are sorted as $\lambda_2< \lambda_3< ...$ such that dominant decay on long time scales in Eq.~\eqref{eq:eigenfunction_expansion} is with the eigenvalue $\lambda_2$. Further, the overlap values $a_{n, \mathrm{ini}}$ are the only pieces in Eq.~\eqref{eq:probability}, that have a dependence on the initial temperature. Therefore $a_{2, \mathrm{ini}}$, which relates to the largest eigenvalue $\lambda_2$, is the relevant quantity when we compare different initial temperatures. The dominant overlap value $a_{2, \mathrm{ini}}$ is  associated to distance measure Eq.~\eqref{eq:dist_measure}; it is proportional to the ordinate intersect of the long time exponential decay in Eq.~\eqref{eq:dist_measure} (see SI), that is $\Delta \mathcal{D} \sim a_{2, \mathrm{ini}}$ (Fig.~\ref{fig:comparison_theory}(b)(inset)). Furthermore, the distance measure and cooling time are proportional with $t_c\sim \mathrm{ln}(\Delta \mathcal{D})$ (see SI).
Using $\Delta \mathcal{D}$ we can now compare the theoretical approach Eq.~\eqref{eq:probability}-\eqref{eq:eigenfunction_expansion} to our numerical simulations of Eq.~\eqref{eq:ABP} (Fig.~\ref{fig:comparison_theory}(a)), showing a good agreement. Here, both approaches show a nonmonotonic behavior of $\Delta \mathcal{D}$ and therefore also in the cooling time, which implies a Mpemba effect.

We now investigate the relevance of activity in our initial distribution $\pi_{\mathrm{ini}}(x)$ and start from an initial distribution without activity ($v_0=0$), which is a Boltzmann distribution. Then at the thermal quench we lower both the temperature of the system and instantaneously turn on the activity of the particle ($v_0\neq0$). The resulting cooling process (Fig.~\ref{fig:comparison_theory}(b)) surprisingly takes longest for $T=T_{\mathrm{cold}}$. Then, at higher temperatures we find two minima, at which $\Delta \mathcal{D} \approx 0$ and therefore the probability distribution Eq.~\eqref{eq:eigenfunction_expansion} decays faster with the smaller eigenvalue $\lambda_3$. In the classification by~\cite{klich2019mpemba} this is referred to as the strong Mpemba effect.

Inspired by the idea that a cooling system subsequently runs through temperatures until it arrives at its steady state, we define a temperature based on the difference measure Eq.~\eqref{eq:dist_measure} that can be traced in time. 
At a given time $t$ we compare the probability distribution $P(x,t)$ to all possible steady states $\pi_{T_{\mathrm{eff}}}(x)$ with effective temperatures $T_{\mathrm{eff}}$, explicitly 
\begin{align}
    \mathcal{D}_{T_{\mathrm{eff}}}(t)  = \sum_i \left| P_i(t) - \pi_{i,T_{\mathrm{eff}}} \right|,
    \label{eq:dist_measure_effT}
\end{align}
which is discretized as before. The temperature $T_{\mathrm{eff}}$ for which $\mathcal{D}_{T_{\mathrm{eff}}}(t)$ is minimal, is then defined as the effective temperature of the system. We now study the evolution of the effective temperature during the cooling process of our model colloid to temperature $T_{\mathrm{cold}}$.
For a passive system the effective temperature decays monotonically, until it reaches its steady state (Fig.~\ref{fig:overcooling}(a)). Turning on activity, we find that the system first goes to  a lower effective temperature than its final steady state temperature (Fig.~\ref{fig:overcooling}(a)). This surprising effect is an \textit{activity induced overcooling} of the system. The probability distributions  (Fig.~\ref{fig:overcooling}(c)) shows how the distribution $P(x,t)$ is closer to an effective distribution $\pi_{T_{\mathrm{eff}}}(x)$ than to the distribution of the cold state $\pi_{\mathrm{cold}}$, meaning that the system has an effective temperature $T_{\mathrm{eff}}< T_{\mathrm{cold}}$. Our theoretical approach using Eq.~\eqref{eq:probability}-\eqref{eq:polarization} also shows an overcooling (Fig.~\ref{fig:overcooling}(a) (inset)). Here, we note that only the long time limit is present due to the truncation of the expansion Eq.~\eqref{eq:eigenfunction_expansion} at the second order. To investigate the dependence of the overcooling effect on the particles' active motion, we compute a phase diagram by varying the self-propulsion and persistence time (Fig.~\ref{fig:overcooling}(d)). At low self-propulsion and persistence we recover the passive case with normal cooling, while increasing both leads to an overcooling.

As an alternative measure for the overcooling of our system we compute the lag diffusion defined as $D(t,\Delta t)= \langle [x(t + \Delta t) -x(t)]^2\rangle / \Delta t$,
where $t$  is the time at which we start measuring the diffusion and $\Delta t$ is the lag time. The lag diffusion can be seen as an alternative dynamical definition of an effective temperature~\cite{szamel2014self}.  By computing the reduced lag diffusion $D(t,\Delta t)-D_{\infty}(\Delta t)$, where $D_{\infty}(\Delta t)$ is the diffusion at the end of our simulations, we observe that the diffusion shows lower values than its final steady state value (Fig.~\ref{fig:overcooling}(e)). This effect is purely induced by activity, it is not present for a passive system (see SI). 

In summary we have shown that a confined colloid induces a Mpemba effect and even a transient overcooling by its activity. We supported our findings  using theoretical calculations revealing a relaxation process that depends on a series of exponential decays. 
In principle our model allows for a direct experimental verification with active colloids.

We have investigated the interplay of two nonequilibrium phenomena, the cooling or heating process of a colloid and its active motion, showing that the active motion of a particle can fundamentally change the relaxation process.
In the future it will be interesting to see how this translates to higher dimensions, and many particle systems. Overcooling enables a fridge to transiently cool a system to lower temperatures than the prescribed temperature which raises the question how the effect optimises the function of coupled heat engines \cite{mamede2021obtaining}.
While we focused on active colloids in this letter, similar effects might arise for biological microswimmers, such as bacteria or microalgae, which can change their motility pattern and therefore their effective temperature in response due to external stimulus~\cite{berg2018random}.

\begin{acknowledgments}
We thank Lorenzo Caprini for insightful discussions. H.L. was supported within the SPP 2065 (project LO 418/25-1).
\end{acknowledgments}

\providecommand{\noopsort}[1]{}\providecommand{\singleletter}[1]{#1}%

\end{document}